\documentstyle[epsf]{elsart}  
\begin{document}
\newcommand{\p}{\partial}
\newcommand{\ls}{\left(}
\newcommand{\rs}{\right)}
\newcommand{\beq}{\begin{equation}}
\newcommand{\eeq}{\end{equation}}
\newcommand{\beqa}{\begin{eqnarray}}
\newcommand{\eeqa}{\end{eqnarray}}
\newcommand{\bdm}{\begin{displaymath}}
\newcommand{\edm}{\end{displaymath}}
%%%%%%%%%%%%%%%%%%%%%%%%%%%%%%%%%%%%%%%%%%%%%%%%%%%%%%%%%%%%%%%%%%%%%%%%%
%                                                                       %
%   BEGIN OF DOCUMENT                                                   %
%                                                                       %
%%%%%%%%%%%%%%%%%%%%%%%%%%%%%%%%%%%%%%%%%%%%%%%%%%%%%%%%%%%%%%%%%%%%%%%%%
\begin{frontmatter}
\title{
Role of the Coulomb interaction in the flow
and the azimuthal distribution of kaons from heavy ion reactions
} 
\author{
Z.S. Wang, Amand Faessler, C. Fuchs, V.S. Uma Maheswari and
D. Kosov
}
\address{Institut f\"ur Theoretische Physik der 
Universit\"at T\"ubingen, D-72076 T\"ubingen, Germany}
%************************************************************************
\begin{abstract}
Coulomb final-state interaction of positive charged kaons
in heavy ion reactions and its impact on the kaon transverse
flow and the kaon azimuthal distribution are investigated within the framework
of QMD ( Quantum Molecular Dynamics ) model. 
The Coulomb interaction is found
to tend to draw the flow of kaons away from that of nucleons and
lead to a more isotropic azimuthal distribution of kaons in the
target rapidity region. The recent FOPI data have been analyzed 
by taking into accout both the Coulomb interaction and
a kaon in-medium potential of the strong interaction. 
It is found that both the calculated
kaon flows with only the Coulomb interaction and with both the Coulomb interaction
and the strong potential agree within the error bars with
the data.
The kaon azimuthal distribution exhibits 
asymmetries of similar magnitude in both theoretical approaches.
This means, the inclusion of the Coulomb potential makes it 
more difficult to extract information of the kaon
mean field potential in nuclear matter from the kaon
flow and azimuthal distribution data. 
\end{abstract}
\begin{keyword}
Kaons, Coulomb interaction, transverse flow, azimuthal distribution, QMD
\\
PACS numbers: {\bf 25.75.+r}
\end{keyword}
\end{frontmatter}
%%%%%%%%%%%%%%%%%%%%%%%%%%%%%%%%%%%%%%%%%%%%%%%%%%%%%%%%%%%%%%%%%%%%%%%%%
%                                                                       %
%   BEGIN OF TEXT                                                       %
%                                                                       %
%%%%%%%%%%%%%%%%%%%%%%%%%%%%%%%%%%%%%%%%%%%%%%%%%%%%%%%%%%%%%%%%%%%%%%%%%
%%%%%%%%%%%%%%%%%%%%%%%%%%%%%%%%%%%%%%%%%%%%%%%%%%%%%%%%%%%%%%%%%%%
\vskip 0.5 true cm
%%%%%%%%%%%%%%%%%%%%%%%%%%%%%%%%%%%%%%%%%%%%%%%%%%%%%%%%%%%%%%%%%%%
\section{Introduction}
%\vskip 0.5 true cm
The importance of Coulomb final-state interaction in particle
production in heavy ion reactions has been 
pointed out more than one decade ago \cite{ben79,bass94,osa96}.
In the energetic collision of two nuclei the Coulomb field 
has been shown to be able to cause a momentum
shift of charged particles and distort the phase space of these particles.
Although both effects are reduced to some extent due to the finite
expansion of the charged sources, they are still very essential in 
understanding the observed $\pi^-$/$\pi^+$ ratios. On the other hand,
production of kaons in heavy ion reactions has attracted 
much attention since it has been found to be sensitive on
the compressibility of nuclear matter \cite{ai85,huang93}.
So it is of high interest to study possible effects of  
final-state Coulomb interaction of charged kaons. In this work we intend to check
the influence of Coulomb interaction on the transverse flow and the azimuthal 
distribution of positive charged kaons. Just recently, G.Q.Li, C.M.Ko and B.A.Li\cite{Li95} indicated that these observables
can act as an unique probe to the kaon mean field potential
in dense nuclear matter. The knowlege about in-medium modifications
of the properties of kaons is important not only to nuclear physics but also
to astrophysics, since it is related to the possible formation
of kaon condensation in the core of neutron stars \cite{kaplan86,brown94}. 
It was mentioned that kaon Coulomb 
interaction has been included in the first calculation of the kaon flow
by Li et al. based on the RBUU (Relativistic Boltzmann-Uhlenbeck-Uehling
) model \cite{Li95}. However, the magnitude and the observable features of the Coulomb effect
are not presented there. Since kaon spectra are caused by
a combination of different physical causes, namely by the in-medium potential
of the strong interaction, by the Coulomb interaction and also the rescattering
of kaons with baryons and pions and by others, we feel it necessary to make clear the
role of the Coulomb final-state interaction in determining the kaon
flow and the kaon azimuthal distribution. The inclusion of these effects
is important to extract
information about the kaon mean field potential in dense matter from the analysis of
the kaon flow and azimuthal distribution.

This paper is organized as following. In section 2 we describe the methods used to treat $K^+$ production
and the Coulomb effect in heavy ion reactions. A main source for $K^+$ production
is the pion-baryon reaction leading to a kaon plus a hyperon. In section 3 we compare
with the experimental data from the FOPI collaboration. In section 4 we summarize the main results of this work.

\section{Description of $K^+$ production and the Coulomb effect}
%\vskip 0.5 true cm
In this work we concentrate on heavy ion reactions at SIS energies 
(about 1-2 GeV/A) since the first experiment of the kaon transverse flow
has been done in this energy region by the FOPI collaboration \cite{ritman95}. 
In order to evaluate the effects of Coulomb interaction
one needs a dynamic model which can describe the charge distribution
in the colliding nuclei as well as the compression and expansion of
the charged sources. In this work we adopt the framework of the QMD (Quantum Molecular Dynamics)
model which meets these requirements. 
It can also distinguish between different isospin
states of nucleons, $\Delta$ and $N^*$ resonances since an effective 
baryon-baryon Coulomb interaction has been implemented in  
QMD \cite{ai86,ai91}. We have extended the model by taking into account
pion-baryon Coulomb interaction \cite{uma97}. This is of importance in providing
a correct pion distribution since the pion phase space has been found
to be highly isospin dependent. The charges of nucleons, nucleonic
resonances (They are mainly $\Delta$ and $N^*$ resonances.) 
and pions are the source of the Coulomb field acting
on $K^+$ mesons at GSI incident energies. The degrees of freedom of these particles
are treated in a non-perturbative way in our QMD model.

\vskip 0.5 true cm
In our model, the $K^+$ mesons are produced via either baryon-baryon
collisions or pion-baryon annihilation processes into kaons and hyperons. For the baryon-baryon 
channel, we have adopted a calculated cross section based on 
the one-boson-exchange (OBE) model \cite{sibi96}, rather than the parametrization
given by Randrup and Ko \cite{ran80}, since a recent experiment has found that
the latter parametrization\cite{ran80} overestimates kaon production near threshold more than 
two orders of magnitude \cite{bal96}. For the pion-baryon channel we have used a cross section calculated
by using a resonance model \cite{tsu95}. This cross section includes the contribution
from relevant resonances up to 2 GeV and agrees well with the existing data. In our previous
work \cite{fu97}, we have shown that the pion-induced channel has a substantial
contribution to kaon production for the incident energy region of
1-2 GeV/A and for large systems. With the inclusion of this production
channel we could reproduce the recent Kaos data for both
the momentum spectra and 
the mass dependence of $K^+$ multiplicities, however, with a 
realistic momentum dependent nuclear mean field. Due to the very
small abundance of kaons in the energy region we have adopted
the standard perturbative method to treat kaon degrees of freedom.
So $K^+$ mesons can not affect the reaction
dynamics. Due to charge conservation, external Coulomb fields have
no effect on the production processes of $K^+$ mesons.
After production of a kaon, it propagates in the Coulomb field
generated by charged nucleons, 
$\Delta$'s, $N^*$ resonances and pions. The Hamiltonian of kaons can be
written as

\beq
H_k = \sum_{\i_k}\left ( \sqrt{m_k^2+{\vec P}_{i_k}^2}+U^{Cou}_{i_k}\right ),
\eeq
where $U^{Cou}_{i_k}$ is the Coulomb potential acting on kaons,
\beq
U^{Cou}_{i_k} = \sum_{i_k}\frac{Z_{i_k}Z_{i_n}e^2}
{\mid{{\vec r}_{i_k}-{\vec r}_{i_n}}\mid}+\sum_{i_r}\frac{Z_{i_k}Z_{i_r}e^2}{\mid{{\vec r}_{i_k}-{\vec r}_{i_r}}\mid}
+\sum_{i_\pi}\frac{Z_{i_k}Z_{i_\pi}e^2}{\mid{{\vec r}_{i_k}-{\vec r}_{i_\pi}}\mid}.
\eeq
Here $i_k$, $i_n$,$i_r$ and $i_\pi$ denote kaons, nucleons, resonances
($\Delta$ and $N^*$) and pions, respectively. The corresponding
charges are $Z_{i_k}$, $Z_{i_n}$, $Z_{i_r}$ and $Z_{i_\pi}$. 
One has to include the Coulomb effects 
from charged nucleonic resonances and
pions at the beam energies of interest, since a substantial number of
nucleonic resonances are created via violent nucleon-nucleon collisions.
The decay of these resonances leads to non negligible abundances of pions. 
However, pion abundances in heavy ion reactions 
are strong isospin dependent. For nuclei with high isospin asymmetry
a large excess of negative charged pions compared to positive charged pions
has been observed. By using the isobar model \cite{sto86}, the $\pi^-$/$\pi^+$ ratio
can be approximately evaluated. For instance, in the case of a reaction system as 
Au + Au one expects a ratio of $\pi^-$/$\pi^+$ as about 1.95 for pions 
from $\Delta$ decay, and 1.75 for pions from $N^*$ decay. 
Since the total charge of the reaction system is conserved
exactly, the baryons are more positive charged during the reaction
than in the ground state of the target and the projectile.
Thus, it is necessary to take into account the contribution
from resonances and pions in order to avoid any overestimation 
of the Coulomb effect. Since it has been shown that the rescattering of
kaons with nucleons can affect the flow and the azimuthal distribution of kaons
too, we have included K-N rescattering in a way similar to Ref.\cite{fang93}.

\vskip 0.5 true cm
Since the Coulomb field on $K^+$ mesons originates from charges of baryons
and pions, we feel it is necessary to make sure that these particles
can be described reliably by our model. In the upper and middle panels of Fig.1 
we present the transverse flow, namely average particle in-plane transverse momenta as a function
of the particle rapidity of nucleons and pions in the reaction of Au+Au at an incident energy
of 1 GeV/A and at an impact parameter of b = 5 fm. The particle rapidities
have been normalized to the projectile rapidity. All through this work,
we have used a soft Skyrme force (corresponding to a nuclear matter
compressibility of K = 200 MeV) combined with a momentum dependent
interaction which was adjusted to the empirical nucleon-nucleon
optical potential. Our results are similar to that based on RBUU by
Li et al.\cite{Li95} for the same reaction in both tendency and magnitude. 
The slight antiflow of pions compared to the nucleon flow is due to reabsorption
of pions by nucleons. 
In the upper and middle panels of Fig.2 we show azimuthal
distributions of nucleons and pions for the same reaction. The distributions
have been normalized to total multiplicities of nucleons and pions, respectively.
We have used a rapidity cut as -1.3$<$$y_{cm}$/$y_{proj}$$<$-0.7 which 
corresponds to the target rapidity region. In this work we place the target nucleus
in the lower and the projectile nucleus in the upper hemisphere.
In Ref.\cite{Li97} it was found that kaon azimuthal distribution in the target (or
projectile) rapidity region is more sensitive to the strong kaon 
potential in dense matter than in the midrapidity 
region. In this rapidity region Coulomb final-state interaction is
also expected
to have a larger effect since the target (or projectile) spectators
have a much lower expansion velocity than the fireball\cite{ben79}. Our results show
that in the target rapidity region the nucleon azimuthal distribution is
peaked at $\phi$=$180^0$ ($\phi$=$0^0$ is defined as the reaction plane in 
the direction of the scattered projectile.)
corresponding to enhanced abundances of nucleons
in the lower hemisphere due to the target spectators. 
This feature is similar to that found in Ref.\cite{Li97}.
For pions a very weak enhancement
in the upper hemisphere can be seen due to the weak antiflow of pions.
In the lower panels of Fig.1 and Fig.2 the kaon
flow and the normalized kaon azimuthal distribution for the same reaction is shown.
The azimuthal distribution is given for the same target 
rapidity region. Both results with and without kaon Coulomb
interaction are shown. The flow and azimuthal distribution
obtained in the case of no Coulomb interaction follow
that of the nucleons, since kaons are produced just from violent 
collision processes in which a nucleon or nucleonic resonance is involved.
We observed a kaon flow in the same direction as that of nucleons. 
The kaon azimuthal distribution in the target rapidity region exhibits
also an enhancement in the lower hemisphere, similar to that of nucleons.
However, the final-state Coulomb interaction distorts
the flow and azimuthal distribution of $K^+$ mesons by a non trivial
magnitude. It can be found from the figures that the 
Coulomb interaction moves the flow of $K^+$ mesons away from that of
nucleons. The preferential emission of target rapidity kaons  
in the lower hemisphere is suppressed by the Coulomb interaction
on $K^+$ mesons. It results in a nearly isotropic azimuthal
distribution of kaons in the target rapidity region. 
In a quantitative language, the Coulomb interaction has changed the kaon
flow parameter, which is defined as F = $\frac{d<P_x>}{dy_{c.m.}}\mid_{y_{c.m.}=0}$,
from 66 MeV/c to 15 MeV/c, approximately. 
Since we have used an azimuthal distribution which is normalized to
the total kaon multiplicity, the quantity 
$\Delta\left (\frac{dN}{d\phi}\right )$
= $(\frac{dN}{d\phi})_{max}$ - $(\frac{dN}{d\phi})_{min}$ can act as a
measure of the asymmetry of the distribution. The asymmetry factor of the kaon
azimuthal distribution decreases from about 0.075 to 0.005.
The influence of the Coulomb
interaction can be understood since positive
charged kaons experience a net repulsive Coulomb field in the reaction
system.

\section{Comparision with data}
%\vskip 0.5 true cm
Now we compare to the experimental data for central collisions of Ni+Ni
at an incident energy of 1.93 GeV/A obtained by the FOPI Collaboration\cite{ritman95}.
We are interested
to check how much freedom is left for the strong kaon potential 
after the inclusion of the Coulomb effect since it also plays
a role in determining the kaon flow and the kaon azimuthal distribution. 
In order to show the contribution from the Coulomb interaction
compared to that of the strong
kaon in-medium potential, we have included the 
propagation of kaons in the strong mean field potential in
our QMD model.
This is also necessary if one expects agreement with the data since
the strong kaon potential has been shown to contribute substantially
to the kaon flow and the kaon azimuthal distribution. The Hamiltonian of kaons
has the form 

\beq
H_k = \sum_{\i_k}\left (\sqrt{m_k^2+{\vec P}_{i_k}^2}+U^{Cou}_{i_k}+
U^{Str}_{i_k}\right ),
\eeq
where $U^{Str}_{i_k}$ is the strong interaction potential
acting on kaons. The long-established
formalism for studying kaon properties in dense matter is
chiral perturbation theory, which treats the chiral symmetry
breaking term in a chiral Lagrangian by current quark masses
as a perturbation since they are very small compared to the nucleon
mass. In this work we have adopted
a potential suggested by Brown and Rho\cite{brown96}.
It includes not only the 
interaction due to vector meson exchange
and the scalar interaction 
originating from the symmetry breaking, but also the scalar
interaction 
arising from virtual pair corrections. This potential takes into account also the change of the 
pion decay
constant in the medium, where it is smaller than in free space due to
a decrease of the quark condensate in nuclear matter. This potential
agrees well to that from the empirical impulse approximation, and
can account for the large attraction for negative charged kaons 
found in kaonic atoms. Some details
of this potential can be found in a recent review given by Brown and Rho\cite{brown96}. 
Since $U^{Str}_{i_k}$ depends not only on the momentum of the kaon and
the surrounding baryon density $\rho_B$, but also on the scalar
density $\rho_S$, we have followed a method used in the RBUU model\cite{web92} to evaluate
the scalar density $\rho_S$ as a function of the baryon density
$\rho_B$. In this work, we have not included the effect of
the strong interaction potential on the kaon production threshold
in elementary collision processes, since there exist quite different
opinions about this point in the literature\cite{fang94,cassing96,brown96,weise93}. 
With respect to enhanced pion
abundances at the higher incident
energies we have included the rescattering of kaons by pions as in Ref.\cite{ko81}. 
On the other hand, kaon-nucleon rescattering is treated in a non-perturbative way
as it was found to be necessary for avoiding possible violation
of energy conservation\cite{Li95}. 
In our calculation, we have used an impact parameter
cut of b$<$4 fm, corresponding to the chosen centrality in the experiment. 
A $P_t$ cut as $P_t$/m$<$0.5
has been used 
in agreement with the experimental acceptance. 
In order to make sure that our QMD model
is reliable at this higher incident energy we compare also the calculated 
proton flow to the data in the upper panel of Fig.3. A good agreement
between the calculation and the FOPI data is found. In the
lower panel of Fig.3 the calculated $K^+$ flow is shown. We have made calculations
for three cases: (1) with neither Coulomb interaction nor
the strong $K^+$ potential, (2) with only the Coulomb interaction, 
(3) with both the Coulomb interaction and the in-medium strong kaon potential.
Although for this light
system the Coulomb interaction has weaker effect compared to the Au+Au system, its contribution
to the kaon flow is visible. They draw the calculated kaon flow closer
to the data. It can be seen that the result with only the Coulomb interaction
lies already whithin the error bars of the data. On the other hand,
the result with both the Coulomb interaction and the strong
interaction potential 
agrees also with the data. It is interesting to note that
if our calculations differ only in the kaon potential of
the strong interaction, both resultant
flows for the kaons of cases (2) and (3) from above are whithin the error bars of the data. It means that the 
experimental precision is not enough to see clearly  the change in
the kaon flow caused by the in-medium potential used here.
In Fig.4 we present the calculated kaon azimuthal distribution
for the same three cases as in Fig.3. 
Here we have adopted the same $P_t$ cut. 
A rapidity cut of
-1.2 $<$ $y_{cm}$/$y_{proj}$ $<$ -0.8 corresponding to the target rapidity
region has been used in getting the distributions. 
It can be found from Fig.3 and Fig.4 that the change in the kaon flow
and azimuthal distribution caused by the Coulomb final-state interaction 
has the same tendency as caused by an in-medium strong kaon potential.
This can be undersood since both the Coulomb interaction and
the strong interaction potential are
effectively repulsive for $K^+$ mesons. We note that the preliminary data
from the FOPI collaboration\cite{ritman95} show no significant asymmetry in the kaon azimuthal
distribution in the target rapidity region.
With neither a Coulomb nor a strong interaction potential the
distribution has a peak around the lower hemisphere, and the corresponding
value of asymmetry factor
$\Delta\left (\frac{dN}{d\phi}\right )$
is about 0.046 . The Coulomb interaction decreases the peak and leads to
a smaller value of
$\Delta\left (\frac{dN}{d\phi}\right )$
as 0.032. The strong potential suppresses
further the emission of kaons in the lower hemisphere and changes
the peak to a dip. However, the resultant asymmetry factor  
$\Delta\left (\frac{dN}{d\phi}\right )$ is about 0.027, which is very
similar to that obtained with only the Coulomb interaction.
These results concerning the flow and the azimuthal distribution of kaons
mean that after taking into account the Coulomb interaction it becomes more
difficult to extract information about the strong kaon potential 
by analyzing the data of the kaon flow and the kaon 
azimuthal distribution.  
Experimental data with higher
precision are needed.

\section{Summary}
%\vskip 0.5 true cm
In summary, we have invesigated the effect of the Coulomb interaction
of $K^+$ mesons with protons, charged resonances and pions
on the kaon flow and the kaon azimuthal distribution. It is found that the final-state
Coulomb interaction can distort the kaon flow and the kaon azimuthal distribution
by a non trivial magnitude. The repulsive Coulomb interaction 
tends to draw the flow of kaons away from that of nucleons and
leads to a more isotropic azimuthal distribution of kaons in the
target rapidity region. These effects are similar in tendency to that  
caused by the strong kaon in-medium potential.
The data obtained recently by the FOPI collaboration have been analyzed
with both the Coulomb interaction and a kaon in-medium potential of 
the strong interaction included. It is found that both the calculated
kaon flow with only the Coulomb interaction and with both the Coulomb interaction
and the strong $K^+$ potential lie within the error bars of the data.
The kaon azimuthal distribution in the same two cases exhibits an 
asymmetry of very similar magnitude, although the peaks in the
two distributions have different positions.
It means, after the Coulomb effect is
taken into account, it becomes more difficult to extract information of the kaon
mean field potential in nuclear matter from the kaon
flow and the kaon azimuthal distribution. 
Higher precision of 
the experimental data is needed to draw a reliable conclusion on the $K^+$
strong potential.
\\
%\begin{ack}
%\vskip 1.0 true cm
%We thank D. Pelte for providing us
%with the FOPI data shown in this work.
%\end{ack}
\newpage
%%%%%%%%%%%%%%%%%%%%%%%%%%%%%%%%%%%%%%%%%%%%%%%%%%%%%%%%%%%%%%%%%%%%%%%%%
%                                                                       %
%   END OF TEXT                                                         %
%                                                                       %
%%%%%%%%%%%%%%%%%%%%%%%%%%%%%%%%%%%%%%%%%%%%%%%%%%%%%%%%%%%%%%%%%%%%%%%%%
%%%%%%%%%%%%%%%%%%%%%%%%%%%%%%%%%%%%%%%%%%%%%%%%%%%%%%%%%%%%%%%%%%%%%

%%%%%%%%%%%%%%%%%%%%%%%%%%%%%%%%%%%%%%%%%%%%%%%%%%%%%%%%%%%%%%
%%%%%%%%%%%%%%%%%%%%%%%%%%%%%%%%%%%%%%%%%%%%%%%%%%%%%%%%%%%%%%%%%%%%%%%%%
%                                                                       %
%   BEGIN OF FIGURES                                                    %
%                                                                       %
%%%%%%%%%%%%%%%%%%%%%%%%%%%%%%%%%%%%%%%%%%%%%%%%%%%%%%%%%%%%%%%%%%%%%%%%%
\newpage
\begin{figure}
\begin{center}
\leavevmode
\epsfxsize = 15cm
\epsffile[0 0 450 420]{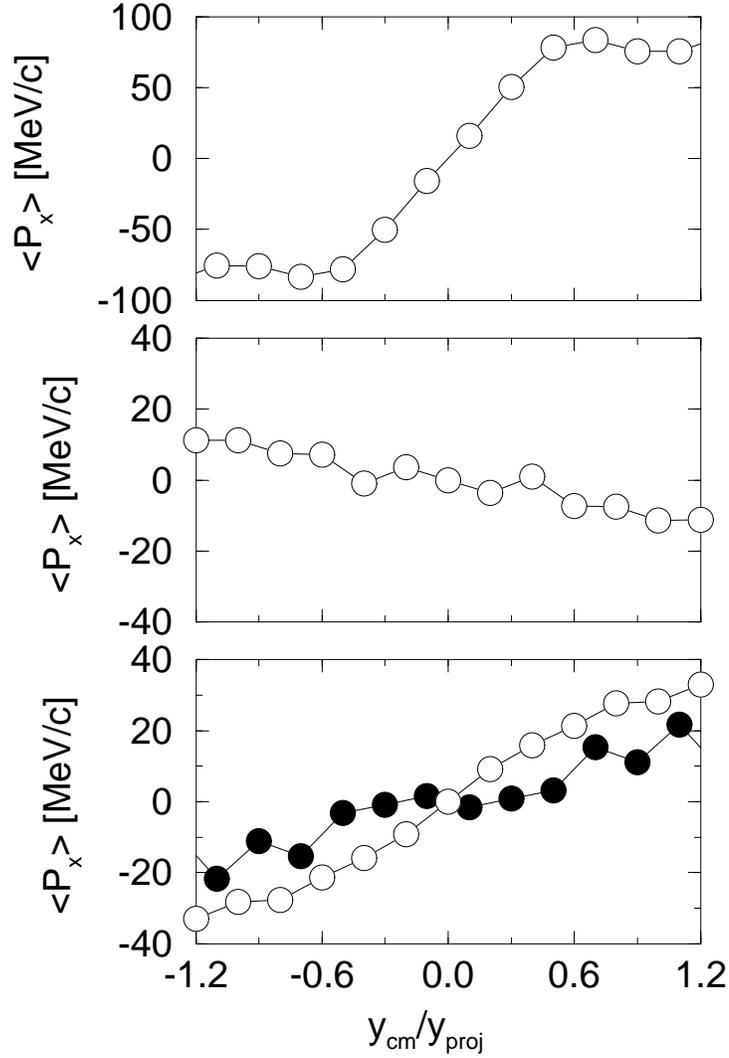}
\end{center}
\caption{
Transverse flows of nucleons (upper panel) and pions (middle panel) as well
as positive charged kaons (lower panel) in the reaction
Au + Au at incident energy E/A = 1 GeV and at impact parameter b = 5 fm.
In the lower panel, the open circles 
denote the calculated kaon flow without any final-state Coulomb
interaction, while the filled circles correspond to the result with
the kaon Coulomb interaction.
}
\label{fig1}
\end{figure}
%%%%%%%%%%%%%%%%%%%%%%%%%%%%%
\newpage
\begin{figure}
\begin{center}
\leavevmode
\epsfxsize = 15cm
\epsffile[0 0 450 420]{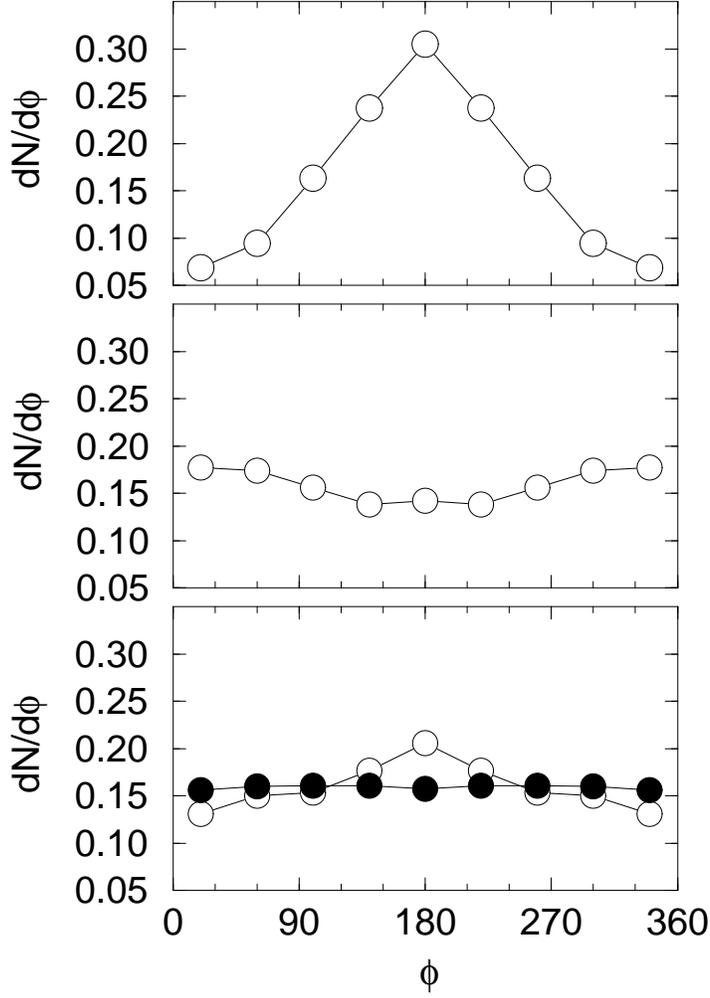}
\end{center}
\caption{
Normalized azimuthal distributions of nucleons (upper panel) and
pions (middle panel) as well as positive charged kaons (lower panel)
 in the reaction of Au + Au at incident energy
E/A = 1 GeV and at impact parameter b = 5 fm. A rapidity cut of
-1.3 $<$ $y_{cm}$/$y_{proj}$ $<$ -0.7 which corresponds to the target rapidity
region has been used in getting the distributions. 
In the lower panel, the open circles 
denote the calculated kaon azimuthal distribution without any final-state Coulomb interaction,
while the filled circles correspond to the result with the 
kaon Coulomb interaction.
}
\label{fig2}
\end{figure}
%%%%%%%%%%%%%%%%%%%%%%%%%%%%%
\newpage
\begin{figure}
\begin{center}
\leavevmode
\epsfxsize = 15cm
\epsffile[0 0 450 420]{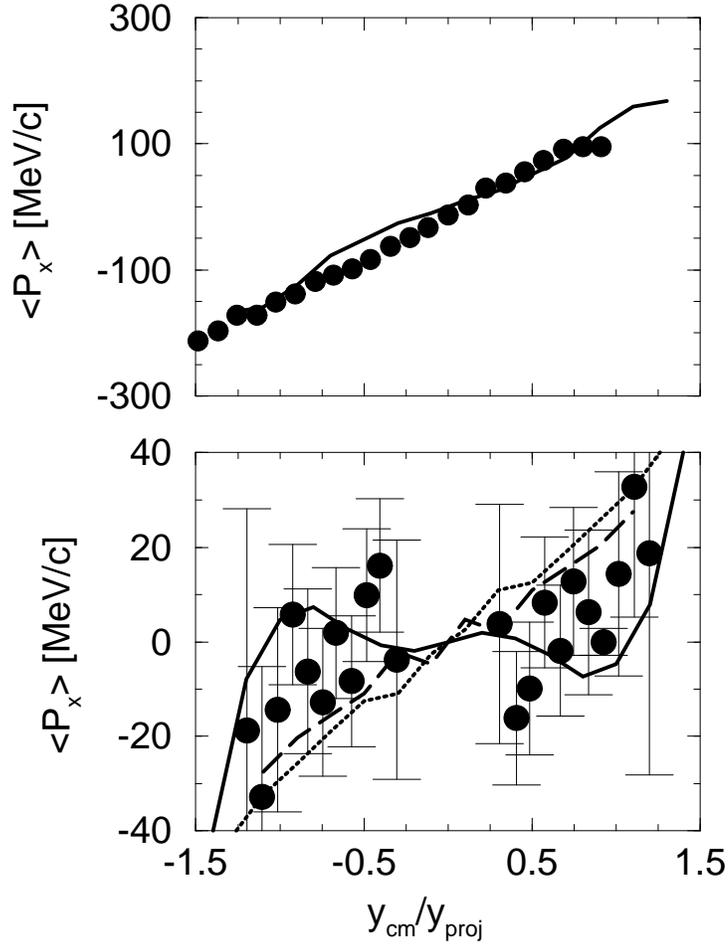}
\end{center}
\caption{
Transverse flows of protons (upper panel) and positive charged kaons (lower panel)
in central collisions of Ni + Ni
at incident energy E/A = 1.93 GeV. The circles are experimental data
obtained by the FOPI group\protect\cite{ritman95}. The lines are our results based on the QMD
model with an impact parameter cut as b$<$4 fm. 
In the lower panel, the dotted line denotes the calculated kaon flow with neither a Coulomb 
nor a strong interaction potential, while the dashed line
denotes the result 
with the Coulomb final-state interaction.
The solid line corresponds to the result with both the Coulomb interaction
and a potential of the strong interaction. A $P_t$ cut as $P_t$/m$<$0.5 has been used
in getting these results.
}
\label{fig3}
\end{figure}
%%%%%%%%%%%%%%%%%%%%%%%%%%%%%
\newpage
\begin{figure}
\begin{center}
\leavevmode
\epsfxsize = 15cm
\epsffile[0 0 450 420]{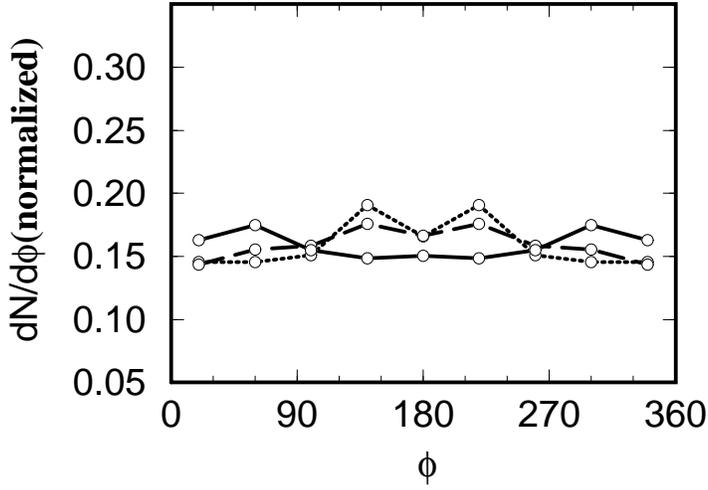}
\end{center}
\caption{
Calculated normalized azimuthal distributions of positive charged kaons
in central collisions of Ni + Ni
at incident energy E/A = 1.93 GeV. 
These results have been obtained based on the QMD
model with an impact parameter cut as b$<$4 fm.
The dotted line denotes the result with neither Coulomb interaction
nor a strong kaon potential, while the dashed line
denotes the result 
with the Coulomb final-state interaction.
The solid line corresponds to the result with both the Coulomb interaction
and a strong kaon potential. 
We have used a $P_t$ cut as $P_t$/m$<$0.5 as in the experiment.
A rapidity cut of
-1.2 $<$ $y_{cm}$/$y_{proj}$ $<$ -0.8 which corresponds to the target rapidity
region has been used in getting the distributions. 
}
\label{fig4}
\end{figure}
%%%%%%%%%%%%%%%%%%%%%%%%%%%%%%%%%%%%%%%%%%%%%%%%%%%%%%%%%%%%%%%%%%%%%%%%%
%                                                                       %
%   END OF FIGURES                                                      %
%                                                                       %
%%%%%%%%%%%%%%%%%%%%%%%%%%%%%%%%%%%%%%%%%%%%%%%%%%%%%%%%%%%%%%%%%%%%%%%%%
%%%%%%%%%%%%%%%%%%%%%%%%%%%%%%%%%%%%%%%%%%%%%%%%%%%%%%%%%%%%%%%%%%%%%%%
%%%%%%%%%%%%%%%%%%%%%%%%%%%%%%%%%%%%%%%%%%%%%%%%%%%%%%%%%%%%%%
\end{document}